\documentclass[pre,groupedaddress,showkeys,twocolumn]{revtex4}
\usepackage{amsmath}
\usepackage{graphics}
\usepackage{graphicx}
\usepackage{amsfonts}
\usepackage{amssymb}
\usepackage{dcolumn}
\usepackage{bm}
\usepackage{natbib}
\usepackage{siunitx}

\usepackage[title]{appendix}%
\begin{document}
\title{Weakly nonlinear creep of amorphous polymers near their glass transition, comparisons between models and experiment}
\author{Martin Roman-Faure}
\author{Antoine Chateauminois}
\email[]{antoine.chateauminois@espci.fr}
\author{Francois Lequeux}
\affiliation{Soft Matter Science and Engineering Laboratory (SIMM), UMR CNRS
7615,
Ecole Sup\'erieure de Physique et Chimie Industrielles (ESPCI), Universit\'e Pierre et Marie Curie, Paris (UPMC), France}
\author{Zhaocheng Zhang}
\author{Catalin Picu}
\affiliation{Department of Mechanical, Aerospace and Nuclear Engineering, Rensselaer Polytechnic Institute, Troy, NY, USA}
%
\begin{abstract}
The nonlinear mechanics of amorphous polymers near the glass transition reveals a stress-induced acceleration of stress relaxation  of nano-metric sub-units. Recent theoretical work predicts that the local acceleration within these nano-domains should scale as the exponential of the squared local stress, a behavior now supported by experiments. However, this local dynamics has some complex consequences on the macroscopic mechanical response, as dynamical heterogeneities generate complex stress and strain fields in polymers close to the glass transition. In this study we consider the nonlinear creep of an amorphous polymer near its glass transition and evaluate the relation between local and global acceleration and the emerging load-carrying structure, by comparing experimental data with predictions of three models of increasing complexity: a two-state (2S) model, a self-consistent (SC) model and a finite-element (FEM) model. The experimentally observed trend of accelerated creep under increasing applied stress is reproduced by the SC and FEM models, while the 2S model overestimates stress localization. The macroscopic, homogenized acceleration is predicted to be close to the microscopic one, albeit with an apparent yield stress that depends on compliance. The FEM model evidences the development of a load carrying substructure that occupies a small fraction of the total material volume driven by the interaction of sub-domains. This work shows that complexity and heterogeneity emerge due to nonlinear interactions and that their adequate representation is essential for predicting the macroscopic mechanical response of amorphous polymers near their glass transition.
\end{abstract}
%
\maketitle
%

%
%
\section{Introduction}
Polymers in the vicinity of their glass transition exhibit large variations of their mechanical properties with temperature. When the temperature is increased by approximately 50~\si{\kelvin}, from below to above the glass transition temperature, $T_g$, the elastic modulus typically decreases from about 1~\si{\giga\pascal} to about 1~\si{\mega\pascal}, while the characteristic relaxation time decreases by approximately three orders of magnitude for every 10~\si{\kelvin}~\cite{ferry_viscoelastic_1980}. Moreover, the relaxation modulus does not decay exponentially but instead follows a stretched exponential, revealing a broad distribution of internal relaxation times. The width of this relaxation-time distribution has been related to the existence of dynamical heterogeneities in glassy polymers~\cite{ediger_spatially_2000}. Accordingly, the material can be viewed as an assembly of nanoscale sub-domains whose relaxation times differ by several orders of magnitude. In this work, we investigate the nature and consequences of this evolving heterogeneity on the macroscopic mechanical behavior of amorphous polymers.

Two limiting cases can be identified depending on whether the experimental time window is much shorter or much longer than the local relaxation-time distribution. These limits correspond to the glassy~\cite{haward_physics_2012} and rubbery states\cite{rubinstein_elasticity_2002}, respectively, both of which are well described in the literature, in the linear and nonlinear regimes. By contrast, the intermediate situation in which the experimental time window overlaps with the relaxation-time distribution of the polymer, i.e., in the vicinity of the glass transition, is considerably richer and more complex because of physical aging~\cite{mckenna_mechanical_2003}. Despite its practical importance, the nonlinear mechanical response in this regime remains only partially understood~\cite{oconnell_large_1997,oconnell_no_2002}.

Various approaches have been proposed to predict the mechanical behavior of amorphous polymers near their glass transition in both the linear and nonlinear regimes~\cite{chen_theory_2011,klochko_stress_2022,caruthers2018quantitative,langer_shear-transformation-zone_2008,merlette_theory_2023}. Building on a physical description of dynamical heterogeneities~\cite{long_heterogeneous_2001,merabia_heterogeneous_2002}, Dequidt~\textit{et al.}~\cite{dequidt_mechanical_2012} modeled the material as an assembly of beads connected by springs with finite lifetimes that evolve as a function of the local stress. In subsequent work, finite element models were developed to predict the linear behavior of glassy polymers represented as statistical assemblies of domains exhibiting dynamical heterogeneities~\cite{masurel_dynamical_2017,montes_modeling_2019}.

The weakly nonlinear mechanical response of glassy polymers was later investigated by Long and coworkers~\cite{dequidt_heterogeneous_2016,conca_acceleration_2017,merlette_eyring_2025}. By weak nonlinearity, we refer to constitutive nonlinearities arising in the small-strain regime, where geometric nonlinearities can be neglected~\cite{roman-faure_weak_2025}. In Refs.~\cite{long_dynamics_2018,merlette_eyring_2025}, it was proposed that the local relaxation time of a sub-domain decreases with increasing local stress $\sigma$ as $e^{-\left(\sigma/y\right)^2}$, where $y$ denotes the local effective yield stress. This prediction contrasts with the widely used Eyring model~\cite{eyring_viscosity_1936}, which predicts a dependence proportional to $e^{-\sigma/y}$. Refs.~\cite{long_dynamics_2018,merlette_eyring_2025} further showed that the quadratic dependence is consistent with optical photobleaching measurements of the segmental dynamics of poly(methyl methacrylate) (PMMA) under applied stress~\cite{lee_direct_2009}, as well as with strain-relaxation~\cite{belguise_weak_2021} and creep experiments~\cite{roman-faure_weak_2025}. Accordingly, the starting point of the present work is the observation that the relaxation time $\tau_i^0$ of a sub-domain is modified by the local stress according to
\begin{equation}
	\tau_i=f\tau_i^0,
	\label{eq:tau}
\end{equation}
where the local acceleration factor is
\begin{equation}
	f=e^{-\left(\sigma/y\right)^2}.
	\label{eq:f}
\end{equation}

The coupling between stress and relaxation time promotes the development of heterogeneity, which may in turn cause the macroscopic acceleration to differ from the microscopic one, as previously demonstrated for stress-relaxation experiments~\cite{belguise_weak_2021,roman-faure_weak_2025}.

The objective of this paper is to clarify the role of the stress-field heterogeneity in determining the macroscopic acceleration during creep by comparing experiments with several modeling approaches. We use creep data recently reported by some of us~\cite{roman-faure_weak_2025}, together with a two-state (2S) model~\cite{roman-faure_weak_2025}, a self-consistent (SC) model, and a finite element model (FEM), to quantify the development of microscopic heterogeneity and its influence on the macroscopic acceleration. We show that the macroscopic acceleration has a functional form similar to that of the microscopic acceleration. The predictions of the SC and FEM models are in good agreement with the experimental observations, whereas the 2S model systematically underestimates the macroscopic acceleration. We further show that a sub-domain substructure that carries the largest fraction of the applied load emerges, evolves toward a steady state, and persists throughout the creep process. These results indicate that field-mediated interactions, captured by the FEM, promote the development of structural complexity, whereas spatial correlations, which are accounted for by the FEM but not by the SC model, are not essential for determining the functional form of the macroscopic response.
%
\section{Experimental data}
\label{sec:experimental}
All creep measurements have been performed using injection molded
poly(etherimide) (PEI) polymer samples (Ultem 1010,
Sabic), an amorphous polymer with glass transition temperature $T_g=213$~\si{\celsius}. Creep experiments were carried out at temperatures ranging from 198~\si{\celsius} to 213~\si{\celsius} and under applied tensile stresses from 1 to 15~\si{\mega\pascal}. In the considered weakly nonlinear domain (see below for a definition of this regime), creep measurements are delicate and require a high accuracy to quantify creep acceleration. For this purpose, both measurements in the linear and  nonlinear regime were carried out while maintaining the specimen in the grips of the tensile machine. A dedicated procedure involving successive thermal annealing and stress steps was developed; the experimental methods and data processing are described in detail in ~\cite{roman-faure_weak_2025}.
\section{Creep modeling}
\label{sec:creep_modeling}
The material in thermodynamic equilibrium is viewed as a collection of sub-domains which are elastically identical, but have different relaxation times, $\tau_i^0$~\cite{montes_modeling_2019,masurel_dynamical_2017,masurel_role_2015}. The relaxation time depends on stress through an acceleration factor $e^{-(\sigma/y)^2}$ where $\sigma$ is the local stress~\cite{conca_acceleration_2017,merlette_eyring_2025} and $y$ a local effective yield stress. This acceleration is the basis of the non linear creep response discussed in this study.\\
It should be mentioned that, at zero stress, the sub-domains' state may evolve under the action of thermal fluctuations. Indeed, an amorphous material below its glass transition and out of thermodynamical equilibrium ages with time with a relaxation time distribution which drifts toward thermal equilibrium. For the PEI material under investigation, we have shown (see supplementary information SI3 in reference~\cite{roman-faure_weak_2025}) that aging affects the measured creep acceleration only at low compliance ($J \lesssim 3. \: 10^{-9}$~\si{\per\pascal})~\cite{roman-faure_weak_2025}. Here, the corresponding experimental data are systematically discarded. Accordingly, our FEM and SC models do not take into account aging and the relaxation times distribution is assumed to remain stationary on the considered time scale.\\
In addition, the relaxation time distribution above the yield stress is known to shift toward shorter times - a phenomenon called rejuvenation~\cite{chen_theory_2011,mckenna_mechanical_2003,nanzai_transition_1990}.  Chain orientation may also modify the dynamics in the yield regime, leading to strain hardening\cite{merlette_theory_2023}. Both effects are beyond the scope of the present work. In order to neglect rejuvenation and chain orientation, as well as geometric nonlinearity, we restrict here our investigations to creep strain less than $\epsilon=0.06$. In reference~\cite{roman-faure_weak_2025}, we showed that below this threshold, creep acceleration for applied stresses up to 15~\si{\mega\pascal} can be accurately described using the Long and coworkers model. Taking into account that $J_0=J(t=0) \approx 6 \;10^{-10}$~\si{\per\pascal}, this strain threshold corresponds to maximum normalized compliance values $J/J_0$ ranging from 100 to 6 when the applied stress was varied from 1 to 15~\si{\mega\pascal}. The corresponding creep times are within the range $t=10-100$~\si{\second}.\\

Within this framework, we consider the SC and FEM models which are both based on the following assumptions: \textit{(i)} each sub-domain in the unstressed state has a specific relaxation time $\tau_i^0$ which is drawn randomly from a lognormal distribution determined experimentally in the linear response regime;and \textit{(ii)} all sub-domains exhibit Maxwell viscoelasticity with the same elastic modulus $G_i=G$ and a relaxation time that depends on the applied local stress $\sigma_i$ according to the recent theoretical prediction by Conca~\textit{et al.}~\cite{conca_acceleration_2017}; thus, $\tau_i=f(\sigma_i)\tau_i^0$. The main difference between the two models is that FEM accounts explicitly for spatial correlations of the stress distribution, which are not accounted for in the mean-field, self-consistent, model.
\section{Two-state model}
\label{sec:2S_model}
The first model considered in the present study, referred to as the 2S model, was introduced in Ref.~\cite{roman-faure_weak_2025}. It provides a lower bound for the predictions of the self-consistent (SC) and finite element (FEM) models. More specifically, two limiting cases can be identified: one in which the stress remains homogeneous throughout the creep process, and another in which the stress is assumed to vanish within relaxed domains. The 2S model corresponds to the latter limit.
In this model, creep is represented as a succession of stress-relaxation events occurring within individual domains, under the assumption that the stress in relaxed domains instantaneously drops to zero and remains constant thereafter. Accordingly, the material is described as an assembly of domains that can exist in one of two states: an unrelaxed elastic state and a relaxed state carrying no stress. Within this framework, creep is governed by the temporal evolution of the relative fractions of relaxed and unrelaxed sub-domains. A detailed description of the model is provided in Appendix~\ref{appendix:appendixA}. As such, the 2S model provides a lower bound for the creep acceleration.
\\
\section{Self consistent modeling}
\label{sec:SC_model}
Here, we introduce a self-consistent (SC) model based on the assumptions outlined in Section~\ref{sec:creep_modeling}. Since the complete derivation is rather technical, we summarize here the main assumptions and governing equations, while deferring the detailed derivation to Appendices~B--D.

We consider the polymer as a heterogeneous material fully tiled by incompressible viscoelastic sub-domains, each described by a Maxwell model. The total strain is assumed to be the sum of elastic and anelastic contributions, an assumption that has proved successful in related problems~\cite{Shi2013,montes_modeling_2019}. For simplicity, we restrict the analysis to the scalar shear components of the stress and strain tensors and assume that the material is incompressible. To obtain an analytical description of the time evolution of the heterogeneous material, mechanical equilibrium is enforced at every time $t$ within a self-consistent framework.

Following the approach developed by Palierne~\cite{palierne_linear_1990}, we begin with Eshelby's solution~\cite{eshelby1957} for an isolated incompressible elastic spherical inclusion embedded in an incompressible elastic matrix. As detailed in Appendix~\ref{appendix:appendixB}, both the stress $\sigma_i$ and the strain $\gamma_i$ are uniform within the inclusion and are related by
\begin{equation}
	\sigma_i=G_m\Gamma+\frac{3}{2}G_m\left(\Gamma-\gamma_i\right),
	\label{eq:sigma_i1_b}
\end{equation}
where $\Gamma$ denotes the applied far-field strain and $G_m$ is the effective shear modulus of the surrounding matrix.

This result can be extended to a material composed of Maxwell-type sub-domains characterized by a relaxation time $\tau_i$, a stress $\sigma_i$, and a strain $\gamma_i$. At time $t$, the strain in each sub-domain is expressed as the sum of the elastic strain stored in the spring, $\gamma_i^e(t)$, and the anelastic strain accumulated in the dashpot, $\gamma_i^a(t)$:
\begin{equation}
	\gamma_i(t)=\gamma_i^e(t)+\gamma_i^a(t),
	\label{eq:maxwell_inclusion_0}
\end{equation}
The constitutive response of the spring is
\begin{equation}
	\sigma_i(t)=G_i\gamma_i^e(t),
	\label{eq:maxwell_inclusion_1}
\end{equation}
whereas that of the dashpot is
\begin{equation}
	\frac{\partial \gamma_i^a}{\partial t}=\frac{\sigma_i(t)}{G_i\tau_i}=\frac{\gamma_i^e(t)}{\tau_i}.
	\label{eq:maxwell_inclusion_2}
\end{equation}

Similarly, the macroscopic strain $\Gamma$ is decomposed into elastic and anelastic components, $\Gamma^e$ and $\Gamma^a$, respectively. The constitutive relation for the effective matrix therefore reads
\begin{equation}
	\Sigma(t)=G_m\left(\Gamma(t)-\Gamma^a(t)\right)=G_m\Gamma^e(t).
	\label{eq:matrix_leastoplastic}
\end{equation}

The self-consistent approximation assumes that the volume averages of the local stress and strain coincide with their macroscopic counterparts, namely
$\Gamma=\sum_i\phi_i\gamma_i$ and $\Sigma=\sum_i\phi_i\sigma_i$, where $\phi_i$ denotes the volume fraction of the $i$th sub-domain. Combining the above equations, as detailed in Appendix~\ref{appendix:appendixC}, yields the evolution equation for the local stress $\sigma_i$ in a sub-domain embedded in a time-evolving matrix of shear modulus $G_m$ and subjected to a macroscopic stress $\Sigma$:
\begin{equation}
	5\frac{\partial \Sigma}{\partial t}-\frac{3G_m+2G_i}{G_i}\frac{\partial \sigma_i}{\partial t}+3G_m\left(\frac{\partial \Gamma^a}{\partial t}-\frac{\sigma_i}{G_i\tau_i}\right)=0.
	\label{eq:time_derivative_b}
\end{equation}

We now specialize to creep under a constant applied stress, such that $d\Sigma/dt=0$. In addition, we assume that all Maxwell sub-domains have the same shear modulus $G_i$, implying $G_i=G_m$. As discussed in Section~\ref{sec:relaxation_time}, the distribution of local relaxation times is obtained by fitting experimental creep data in the linear regime. Enforcing the self-consistency condition (see Appendix~\ref{appendix:appendixD}) allows Eq.~(\ref{eq:time_derivative_b}) to be recast in the compact matrix form
\begin{equation}
	\frac{\partial \sigma_i}{\partial t}=M_{ij}\sigma_j,
	\label{eq:system_b}
\end{equation}
where
\begin{equation}
	M_{ij}=\frac{3}{5}\left(\frac{\phi_j}{\tau_j}-\frac{\delta_{ij}}{\tau_i}\right),
	\label{eq:matrixM1_b}
\end{equation}
with $\delta_{ij}$ denoting the Kronecker delta. The initial condition is $\sigma_i(t=0)=\Sigma$. The resulting system is solved numerically (see Appendix~\ref{appendix:appendixE}) to obtain the time evolution of the local stress $\sigma_i$ together with the elastic and anelastic strains, $\gamma_i^e$ and $\gamma_i^a$, for a prescribed relaxation-time distribution. In the nonlinear regime, the relaxation time $\tau_i$ depends on the local stress $\sigma_i$, as described in Section~\ref{sec:relaxation_time}. Consequently, the local relaxation times are updated at every integration step.\\
%
%
\section{FEM Modeling}
\label{sec:FEM_model}
Finite element models representing material domains partitioned into sub-domains with different relaxation times are developed to explore the relation between material scale creep and the microscale relaxation times distribution. Cubic domains of edge length $L$ divided into $16^3$ sub-domains are considered. Each subdomain is discretized in $3^3$ linear brick elements (C3D8 in Abaqus), each element having edge length $L/48$ in the undeformed configuration. \\
All sub-domains have the same shear and bulk moduli, $G_i$ and $K_i$, while time-dependent behavior is allowed only in the shear mode. The relaxation time for sub-domains, $\tau_i$, is sampled from a distribution, as described in section~\ref{sec:relaxation_time}. A Poisson ratio of 0.49 is selected to ensure close to incompressible deformation.\\
Uniaxial tension is applied by specifying a macroscopic normal tensile stress $\Sigma_{1 1}$ acting in the $x_1$ direction (referred to below as $\Sigma$, for simplicity). All other macroscopic stress components are kept zero. Periodic boundary conditions are applied in all three directions. Rigid body translations are removed by restricting the displacements of one of the model nodes. The quantity of interest in creep is the variation in time of the stretch in the $x_1$ direction. The solution is obtained with the finite element solver Abaqus, version 2022. 
%
%
\section{Relaxation time distribution}
\label{sec:relaxation_time}
\subsection{Linear regime}
\begin{figure} [!ht]
	\centering
	\includegraphics[width=1\linewidth]{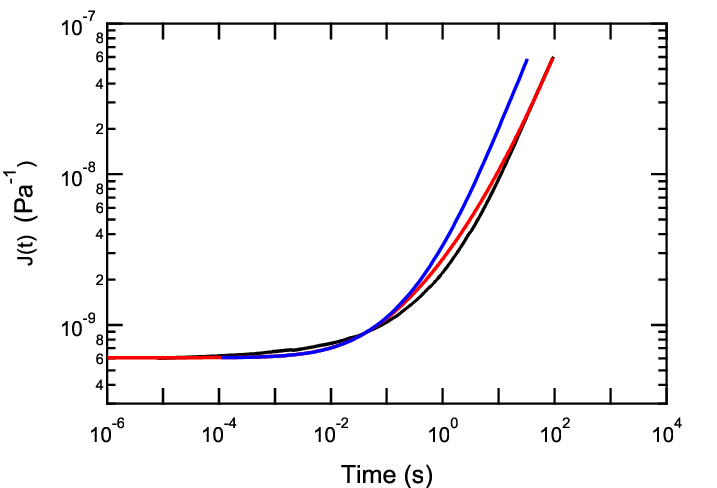}   
	\caption{Experimental (black) and simulated (blue: FEM; red: SC model) creep compliance $J(t)$ of PEI as a function of time in the linear regime at $T_{ref}=T_g=213$~\si{\celsius}. All curves correspond to $\Sigma=1$~\si{\mega\pascal}.  $J(t)$ predicted by the SC model was fitted to the experimental data by considering a lognormal probability distribution of the relaxation time with $s=1.86 \pm 0.03$ and $\tau_0=0.21 \pm 0.01$~\si{seconds} and taking $J_0=6.06\;10^{-10}$~\si{\per\pascal}. The FEM curve is predicted by using a constitutive model of same parameters.}
	\label{fig:J_t_lin}
\end{figure}

In the FEM and SC models we consider identical relaxation time distributions assumed to be of lognormal type as in Refs.~\cite{masurel_role_2015,montes_modeling_2019}:
\begin{equation}
	P(\ln(\tau_i))=\frac{1}{\sqrt{2\pi s}}\exp\left(-\frac{\ln\left(\frac{\tau_i}{\tau_0}\right)^2}{2s^2}\right).
\end{equation}
The mean $\tau_0$ of the lognormal distribution and its width $s$ are obtained from a least square-fit of the SC model to the experimental creep compliance master curve of an amorphous PEI polymer (using $T_{ref}=T_g=213$~\si{\celsius} as the reference temperature). Full details regarding the determination of the experimental master curve are provided in reference~\cite{roman-faure_weak_2025}. Using the numerical scheme described in Appendix~\ref{appendix:appendixE} for the calculation of $J(t)$, the least-square fit to experimental data provide $s=1.86 \pm 0.03$ and $\tau_0=0.21\pm 0.03$~\si{seconds} with evenly distributed relaxation times on a logarithmic scale from $10^{-7}$ to $10^{6}$~\si{\second}. Here, $\tau_0$ is associated with the glass transition, while $s$ characterizes the width of the relaxation time distribution. A comparison between experimental and simulated creep compliance curves in the linear regime is provided in Fig.~\ref{fig:J_t_lin}. The SC model with Maxwell elements allows a reasonable description of the macroscopic linear creep behavior within the considered time range. The ability of the SC model to describe the frequency dependence of the linear viscoelastic modulus in the glass transition range is also detailed in
Supplementary Information SI1.\\
Using the same relaxation time distribution in the FEM model leads to a predicted creep compliance $J(t)$ that is shifted to shorter times as compared to the SC model. This acceleration of creep kinetics may be attributed to spatial correlations which are not accounted for in the mean-field SC model~\cite{masurel_role_2015}.
\subsection{Nonlinear regime}
As per the physical assumptions of the model, in the weakly non linear regime, stress is assumed to induce a shift of the relaxation time distribution to shorter times in the absence of any geometric nonlinearities, according to Eqs.~(\ref{eq:tau}) and (\ref{eq:f}). The modeling approaches thus require the local effective yield stress $y$ in Eq.~(\ref{eq:f}), but creep experiments only yield a macroscopic stress-induced acceleration factor $F$ (see next section for details) which was found in reference~\cite{roman-faure_weak_2025} to be well parametrized by the expression
\begin{equation}
	F = e^{-\left(\Sigma/Y\right)^{2}},
	\label{eq:F_macro}
\end{equation}
where the macroscopic effective yield stress $Y$ is observed to depend on compliance $J$ due to the development of stress inhomogeneities during the course of creep deformation. By definition, however, $y$ is independent of $J$. Therefore, $y$ is identified with $y = Y (J(t \rightarrow 0))$, as stress may be assumed to be homogeneous (i.e. $\sigma_i = \Sigma$) in the limit $t \rightarrow 0$. For the present system, we obtained $y=14$~\si{\mega\pascal}~\cite{roman-faure_weak_2025}. Note that at this stage, all parameters of the SC and FEM are fully defined.
%

%
\section{Nonlinear creep compliance and acceleration factor}
\begin{figure} [!htp]
	\centering
	\includegraphics[width=1\linewidth]{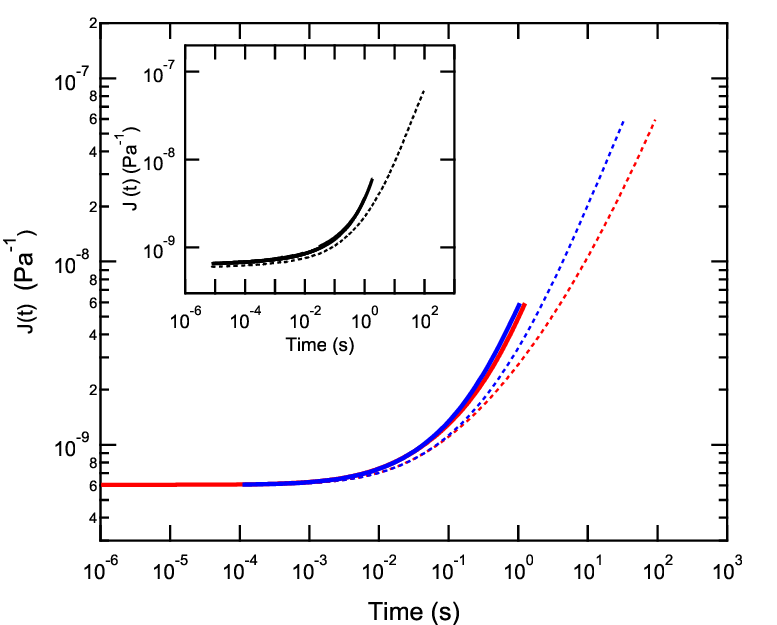}  
	\caption{Calculated creep compliance $J(t)$ as a function of time in the linear (dotted lines) and non linear (solid lines; $\Sigma=10$~\si{\mega\pascal} ) regimes. Red: SC model; blue: FEM simulations. Inset: experimental creep compliance curves of PEI for $\Sigma=$1 (dotted line) and $\Sigma=$10~\si{\mega\pascal} (solid line) at $T_{ref}=T_g=213$~\si{\celsius} (data taken from reference~\cite{roman-faure_weak_2025}).}
	\label{fig:Jt_non_linear}
\end{figure}

\begin{figure} [!htp]
	\centering
	\includegraphics[width=0.8\linewidth]{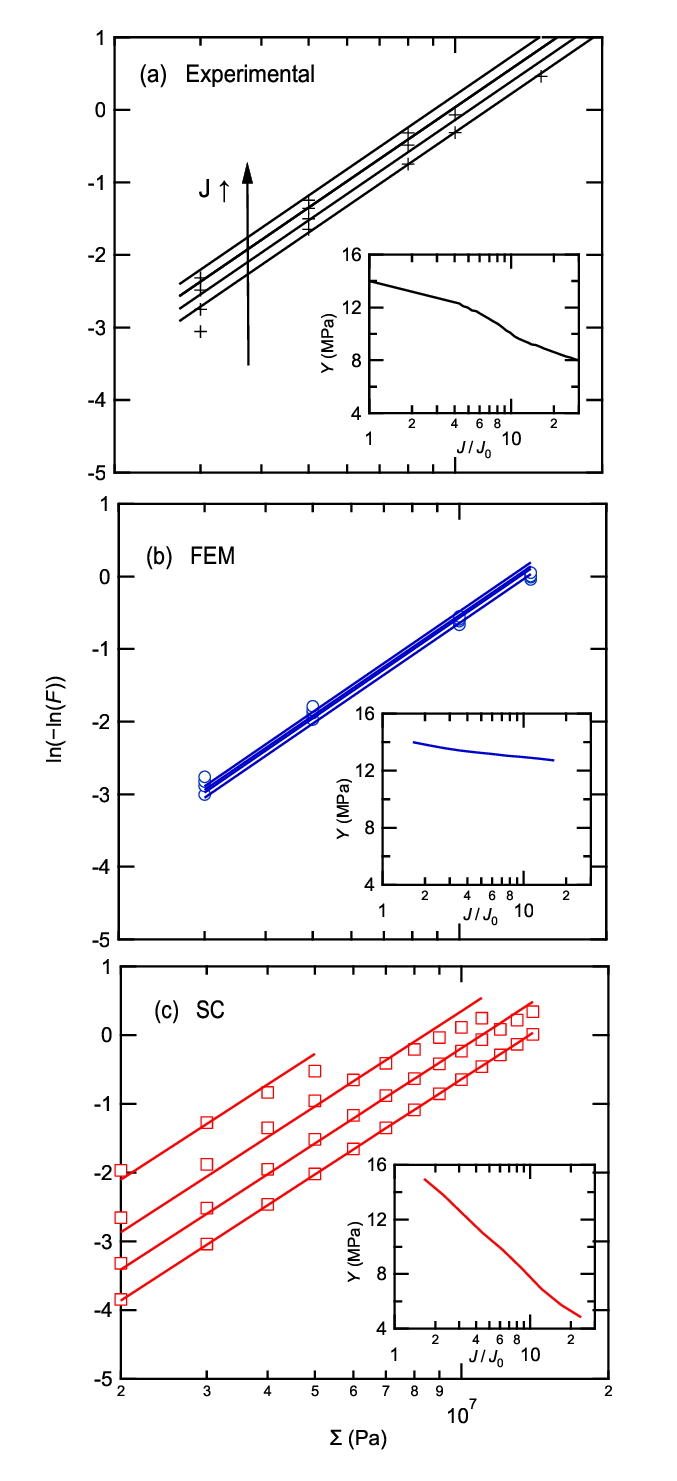}     
	\caption{$\ln(-\ln{F})$, where $F$ is the macroscopic acceleration function, as a function of the logarithm of the applied stress $\Sigma$ for increasing values of the normalized compliance $J(t)/J_0$. (a) experimental results taken from \cite{roman-faure_weak_2025} with, from bottom to top: $J(t)/J_0=$~5, 8.3, 11.7, 16.7; (b) FEM simulations with $J/J_0=$~2.12, 4.58, 7.64 and 16.87; (c) SC simulations with $J(t)/J_0=$~2.30, 4.48, 8.73, 19.98. Simulations were carried out with $s=1.86$. The solid lines are linear fits of the data with a slope equal to 2. Insets show the changes in the parameter $Y$ in Eq.~(\ref{eq:F_macro}) as a function of the normalized compliance $J/J_0$.}
	\label{fig:lnf_lns}
\end{figure}

\begin{figure} [!htp]
	\centering
	\includegraphics[width=1\linewidth]{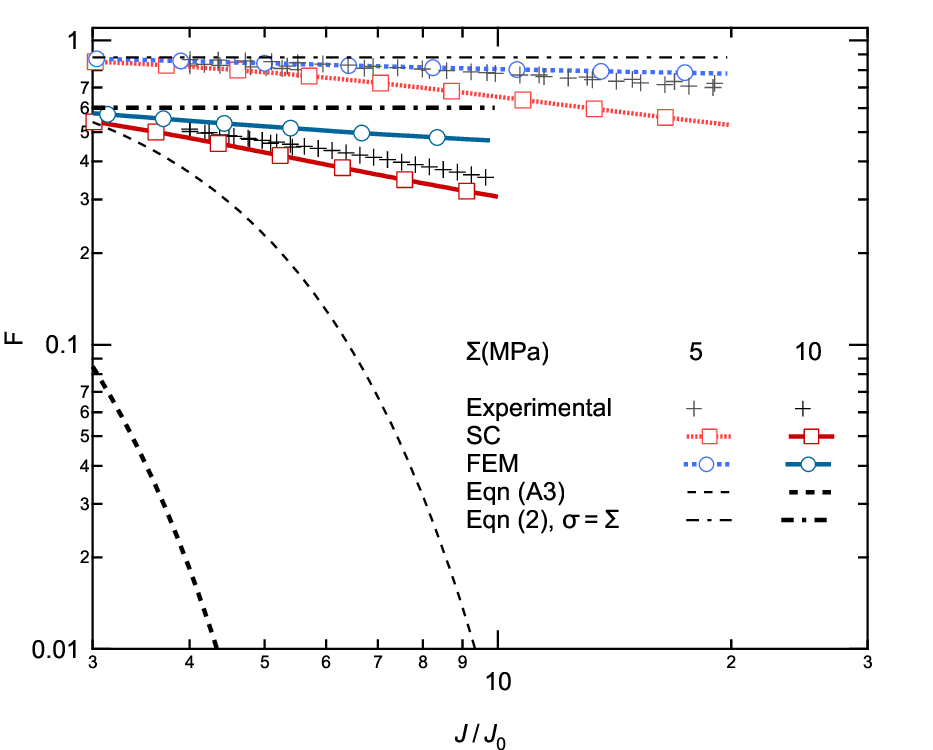}    
	\caption{Log-Log representation of the macroscopic acceleration function $F=t_{NL}/t_{L}$ versus the normalized creep compliance $J(t)/J_0$ for $\Sigma=$~5 and 10~\si{\mega\pascal} ($s=1.86$). Dashed-dot and dashed lines provide upper and lower bounds for the simulated and experimental values which are derived, respectively, for a system with an homogeneous stress distribution (Eq.~(\ref{eq:f}) with $\sigma=\Sigma$) and from the 2S model (Eq~\ref{eq:Fcav}).}
	\label{fig:F_J}
\end{figure}

Continuous lines in Figure~\ref{fig:Jt_non_linear} show the calculated creep compliance curves for an applied stress in the non linear regime ($\Sigma=10$~\si{\mega\pascal}). As a reference, $J(t)$ curves obtained in the linear regime are also shown as dotted lines. Both FEM and SC calculations show creep acceleration qualitatively consistent with the experimental $J(t)$ master curves for PEI, which are reported in the inset for $\Sigma=1$~\si{\mega\pascal} (\textit{i.e. }in the linear regime) and $\Sigma=10$~\si{\mega\pascal}.\\
%
In order to quantify the time-shift of the creep compliance in the non linear regime, a macroscopic acceleration function is defined as $F(J,\Sigma)=t_{NL}/t_L$ where $t_{L}$ and $t_{NL}$ are respectively the times required to achieve a given compliance level $J$ in the linear and nonlinear regimes under an applied stress $\Sigma$.\\
We focus next on the analysis of the stress dependence of the acceleration factor $F$.
In Fig.~\ref{fig:lnf_lns}, $\ln(-\ln F)$ is plotted  as a function of $\ln\Sigma$ for increasing values of the normalized compliance $J(t)/J_0$, where $J_0=J(t=0)$. Fig.~\ref{fig:lnf_lns}(a,b,c) refers to experimental data, FEM simulations and SC calculations, respectively. In accordance with Eq.~(\ref{eq:F_macro}), a linear relationship with a slope close to $2$ is obtained with both SC and FEM models as well as with the experimental results. Note that this property is also verified for the two-states model, but not in the case of step-strain relaxation experiments \cite{belguise_weak_2021} where an exponent of $0.83$ was observed experimentally.\\
in Fig.~\ref{fig:lnf_lns}, a shift to high $\ln(-\ln F)$ values is also observed when $J$ increases. Taking into account Eq.~(\ref{eq:F_macro}), this shift corresponds to a decrease in $Y$ with increasing $J(t)$ which is shown in the inset to Figs.~\ref{fig:lnf_lns}(a,b,c). The models are thus in good qualitative agreement with the experimental behavior. Indeed, the self-consistent model overestimates the decreases of $Y$ with increasing $J$, while the FEM model underestimates it.\\
%
Another way to analyze creep acceleration is to plot the variation of $F$ with the normalized creep compliance $J(t)/J_0$. Experimental $F$ vs. $J(t)/J_0$ data from reference~\cite{roman-faure_weak_2025} for applied stresses of $\Sigma=5$ and 10~\si{\mega\pascal} are shown in Fig.~\ref{fig:F_J}, together with FEM and SC predictions. It turns out that both FEM and SC models capture the experimental trends, \textit{i.e.} a decrease in $F$ with increasing $J$ for a given applied stress, without any adjustable parameters. The FEM model slightly overestimates $F$ values, while the SC model underestimates them. This shows the effect of field-mediated spatial correlations of sub-domains on the acceleration function; specifically, correlations increase the acceleration. The 2S model, which assumes that stress is carried only by the sub-domains that did not relax, predicts a rapid decay of $F$ with increasing $J$, which becomes very pronounced at higher stresses (see dotted lines in Fig.~\ref{fig:F_J}). This drastic departure from the experimental data indicates that load redistribution is critical during creep. In the other limit, assuming a homogeneous and time-independent stress, i.e. $\sigma=\Sigma$, results in a constant acceleration factor $F=f$, as shown by the dash-dotted lines in Fig.~\ref{fig:F_J}. This overestimation is expected, considering that stress localization is a key component of this physics.\\
We also considered the effect of the width of the relaxation time distribution on the acceleration factor. As detailed in Supplementary Information SI2, increasing the distribution width result in a more rapid decrease of $F$ as $J$ increases. In other words, creep acceleration is enhanced by material disorder.\\ 

Some comments are in order regarding the values of the macroscopic acceleration exponent obtained under creep and step-strain loadings. Surprisingly, the macroscopic exponent observed under creep loading is remarkably robust: it is the same for all models and, therefore, independent of the degree of stress disorder. Furthermore, this macroscopic exponent is identical to the microscopic one. The situation is very different for stress relaxation experiments \cite{belguise_weak_2021}. In such experiments, the acceleration exponent can only be measured over a time range during which the macroscopic stress typically varies by a factor of $5$. The resulting macroscopic exponent is about $0.8$, whereas the microscopic exponent is $2$. Numerical simulations similar to those presented here reproduce this experimental value and attribute it to the development of stress inhomogeneities. Creep is therefore a special case in which the macroscopic acceleration exponent is practically equal to the microscopic exponent, most likely because the average stress remains constant.\\
%
%
\section{Local stress distribution}
\begin{figure} [!ht]
	\centering
	\includegraphics[width=1\linewidth]{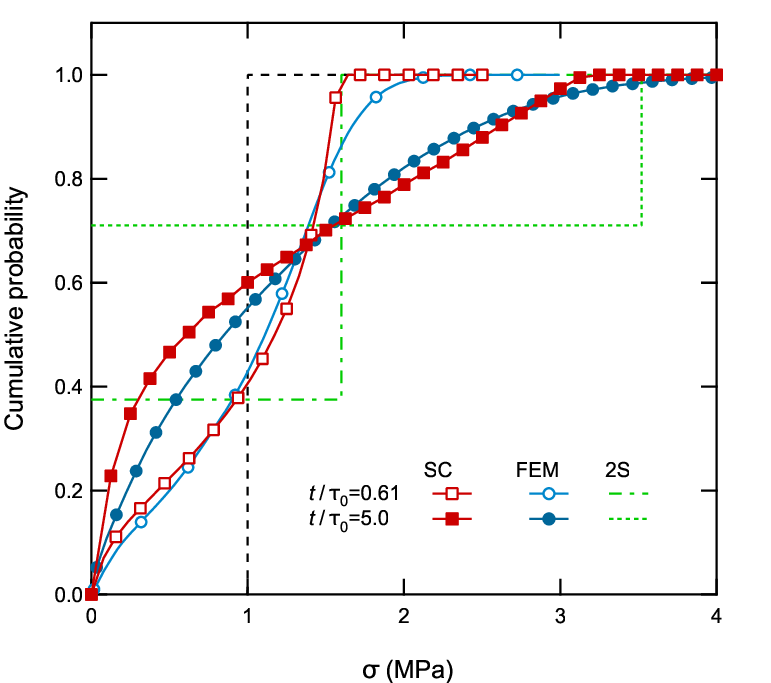}    
	\caption{Cumulative probability distribution of the local stress $\sigma$ for different creep times in the linear regime ($\Sigma=1$~\si{\mega\pascal}, $s=1.86$). FEM simulations refers to von Mises stress. At $t/\tau_0=0.61$ and $t/\tau_0=5.0$, the normalized compliance is $J/J_0=2 \pm 0.02$ and $J/J_0=5.2 \pm 0.6$, respectively, for both FEM and SC simulations. The green dotted and dash-dotted lines are the prediction of the 2S model at the corresponding $J/J_0$ values. The black dashed line indicates the distribution at the beginning of creep ($t/\tau_0=5.23\;10^{-4}$).}
	\label{fig:distri_sigma_lin}
\end{figure}

SC and FEM models provide the stress distribution within the system. Fig. \ref{fig:distri_sigma_lin} shows the cumulative probabilities of stress distribution which are predicted by the two models in the linear regime ($\Sigma=1$~\si{\mega\pascal}) for $t/\tau_0=0.61$ and 5.0. At these creep times, nearly identical values of the compliance are achieved: $J/J_0=2 \pm 0.02$ at $t=0.61$ s and 
$J/J_0=5.2 \pm 0.6$ at $t=5$ s in both SC and FEM. The black dashed line in Fig.~\ref{fig:distri_sigma_lin} shows the cumulative distribution corresponding to a delta function, which represents the homogeneous stress distribution at the onset of creep ($t/\tau_0=5.23\;10^{-4}$). The distributions predicted by the 2S model are also included (dashed and dashed-dot green lines); they consist of two stress levels - one at zero and the other at the average stress of the unrelaxed sub-domains (given by Eq.~(\ref{eq:sigma_two-states})). The FEM and SC models provide similar predictions: a distribution that smoothly increases from $0$ to $1$ over a stress range that increases with time, while the 2S model predicts an increase in the fraction of unloaded sub-domains and a proportional increase in the stress carried by the unrelaxed sub-domains. The broadening of distributions indicates developing heterogeneity.\\

\begin{figure} [!ht]
	\centering
	\includegraphics[width=1\linewidth]{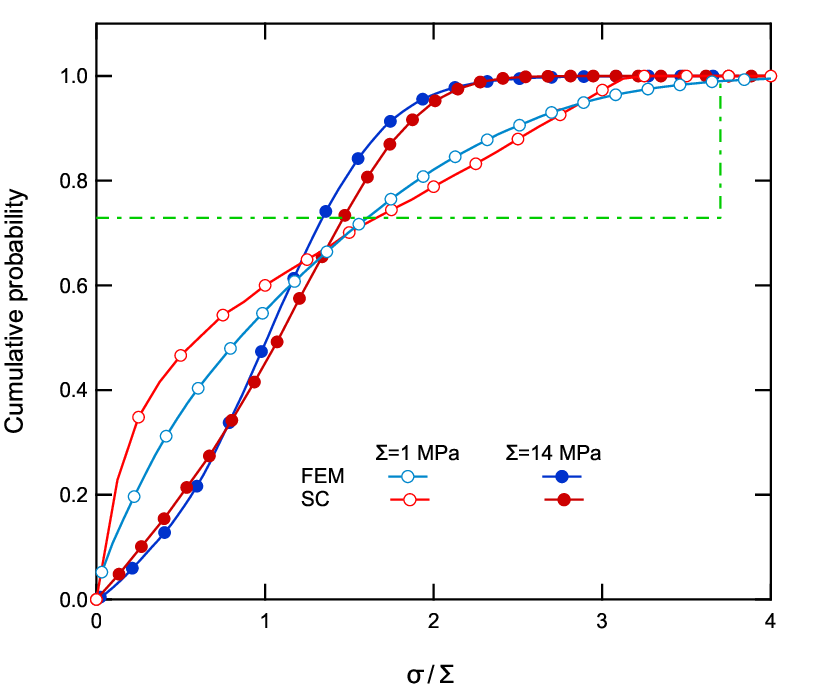}   
	\caption{Cumulative probability distribution of the local stress in the nonlinear regime at $\Sigma=14$~\si{\mega\pascal} and $t/\tau_0=1.43$ predicted by the SC and FEM models. The normalized compliance is $J/J_0=5.3$ and 5.7 for the SC and FEM simulations. Data in the linear regime ($\Sigma=1$~\si{\mega\pascal}) for similar compliance ($J/J_0=5.2$, taken from Fig.~\ref{fig:distri_sigma_lin}) are also reported for reference. FEM simulations refers to von Mises stress. The green dotted line is the prediction of the 2S model at  for $J/J_0=$~5.5.}
	\label{fig:distri_sigma_nonlin}
\end{figure}

\begin{figure} [!ht]
	\centering
	\includegraphics[width=1\linewidth]{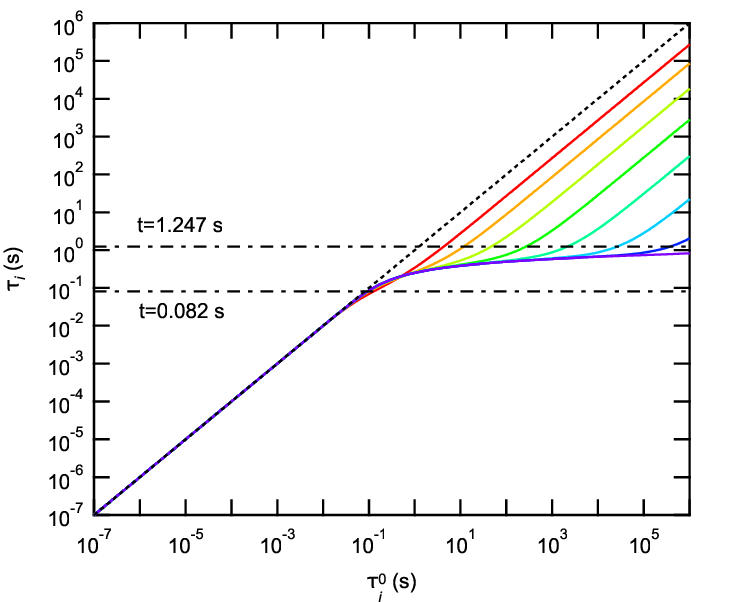}   
	\caption{Relaxation time during creep in the nonlinear regime as a function of the initial relaxation times $\tau_i^0$ ($\Sigma=10$~\si{\mega\pascal}, SC model). From red to purple line, the normalized compliance  $J/J_0$ increases from 2 to 10 by unity. The dotted line corresponds to an affine distribution $\tau_i=\tau_i^0$. Horizontal dash-dotted lines correspond to creep times $t=0.082$~\si{\second} ($J/J_0=2$) and $t=1.2477$~\si{\second} ($J/J_0=10$), respectively.}
	\label{fig:taui_vs_taui0}
\end{figure}

Figure~\ref{fig:distri_sigma_nonlin} shows the stress distribution in the nonlinear regime ($\Sigma=14$~\si{\mega\pascal}) at $t/\tau_0=1.43$. Here again, similar values of the compliance are reached at this creep time in the SC ($J/J_0=5.3$) and FEM ($J/J_0=5.7$) simulations.  The distributions in the linear regime with $J/J_0=5.2$ (data taken from Fig.~\ref{fig:distri_sigma_lin}) are also reported in the figure for reference. In the nonlinear regime, the cumulative probability distribution of stress becomes steeper than in the linear regime.
%
The applied stress forces the relaxation of the domains that have initially (or in the linear regime) a relaxation time greater than the elapsed time. This is shown in Fig.~\ref{fig:taui_vs_taui0} where the relaxation times $\tau_i$ in the stressed state are shown as a function of their values $\tau_i^0$ in the unstressed state for increasing compliance in the nonlinear regime ($\Sigma=10$~\si{\mega\pascal}). In this figure, the horizontal dash-dot lines delimit the considered range of creep times. It emerges that creep is progressively depopulating the part of the distribution with initial relaxation times $\tau_i^0$ higher than the creep time.

At the same time, the stress distribution broadens, as illustrated in Figs.~\ref{fig:distri_sigma_lin} and \ref{fig:distri_sigma_nonlin}. Despite this broadening, the acceleration measured at a fixed value of $J$ remains well described by the expression $e^{-\left(\Sigma/Y(J)\right)^2}$. This indicates that the apparent yield stress $Y$ is unaffected by the applied macroscopic stress and, consequently, by the changes in the stress distribution induced by nonlinear effects.\\

\begin{figure} 
	\centering
	\includegraphics[width=1\linewidth]{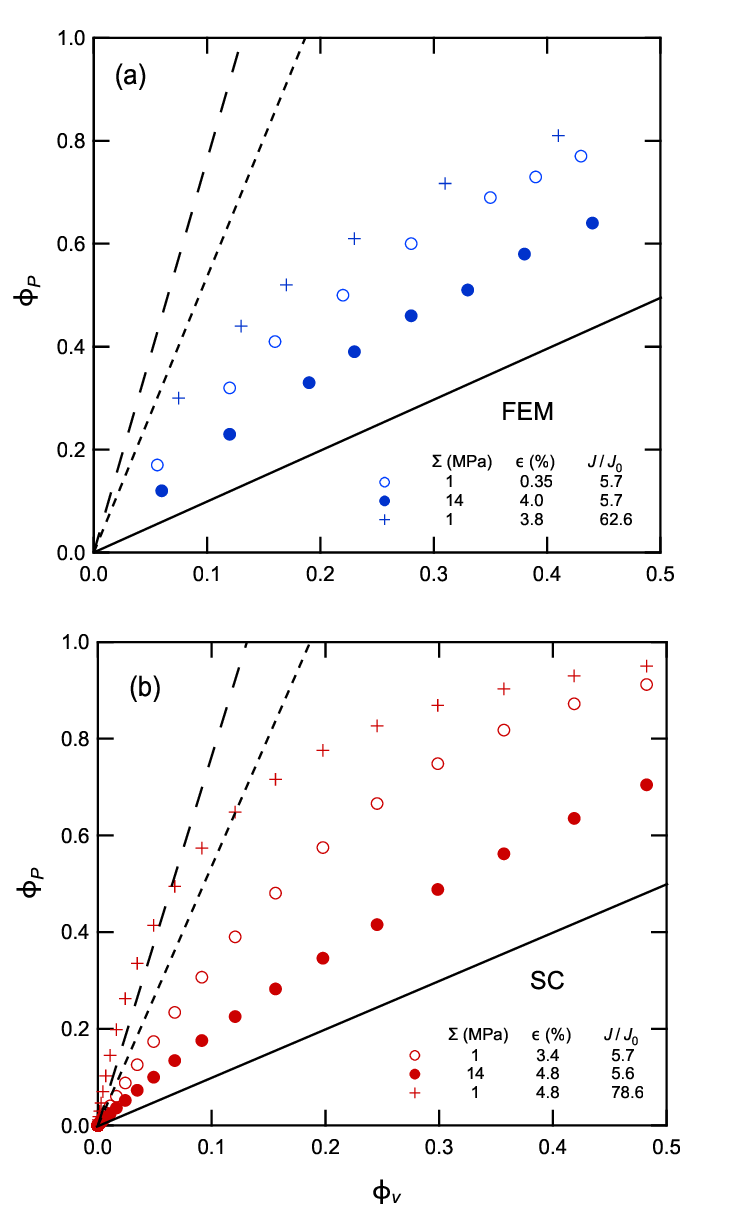}  
	\caption{Fraction $\phi_P$ of the total load carried by the most loaded elements as a function of their volume fraction  $\phi_v$ for (a) FEM and (b) SC models for two applied stress levels $\Sigma$=1 and 14~\si{\mega\pascal} and for different levels of normalized compliance $J/J_0$ and creep strains ($s=$~1.86). Short and long dashed lines correspond to the prediction of the 2S model (Eq.~(\ref{eq:psi})) for $J/J_0=$~3.8 and 5.7, respectively. The solid line is the affine prediction.}
	\label{fig:fraction_load}
\end{figure}

\begin{figure} [!ht]
	\centering
	\includegraphics[width=1\linewidth]{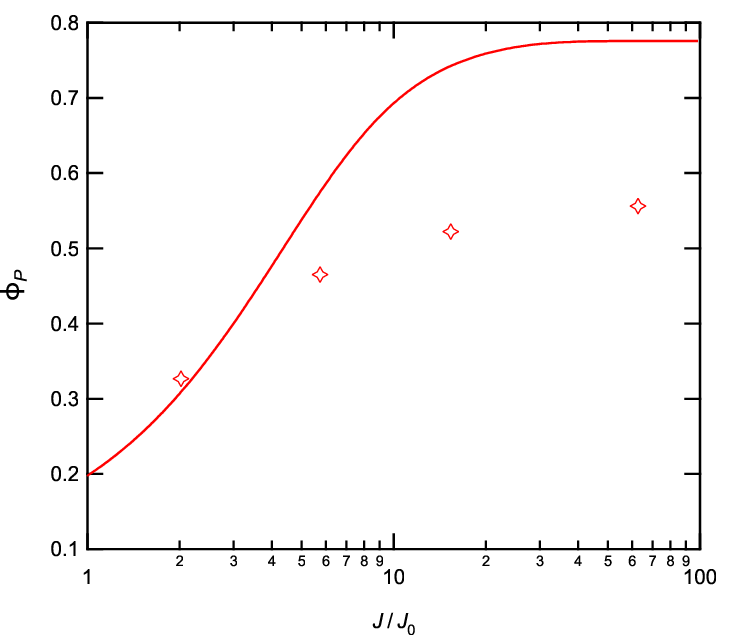}   
	\caption{Fraction $\phi_P$ of the total load carried by the most loaded sub-domains of total volume fraction $\phi_v=20$~\% as a function of the normalized compliance $J/J_0$ for $\Sigma=14$~\si{\mega\pascal}. Solid line and symbols correspond to SC and FEM models, respectively.}
	\label{fig:fraction_total_load_phi0p2}
\end{figure}

A complementary picture of stress distribution may be obtained by plotting the fraction of the total load, $P$, carried by the most loaded sub-domains, $\phi_P=\frac{\int_\sigma^{\infty}A_d p(\sigma)\sigma d\sigma}{P}$, \textit{vs.} the volume fraction of the respective domains, $\phi_v$. $A_d$ is the area of a sub-domain normal to the applied load. Figure~\ref{fig:fraction_load} shows $\phi_P(\phi_v)$ predicted by the FEM and SC models, in the linear, $\Sigma=1$, and nonlinear, $\Sigma=14$, regimes for two values of the normalized compliance, $J/J_0$. Predictions of the 2S model corresponding to the same normalized compliance are included. The continuous line indicates the affine limit which corresponds to a uniform (homogeneous) stress state. It is seen that stress heterogeneity is present at all stress levels and all times: a small fraction of the sub-domains carry a large fraction of the applied (time-invariant) load. Comparing data for given $\Sigma$ and increasing $J/J_0$, it is observed that the degree of heterogeneity increases. Further, comparing the data corresponding to same $J/J_0$ and increasing $\Sigma$, the degree of heterogeneity decreases, which indicates that increasing the applied stress renders the microstructure more uniform, i.e. it has a smoothing effect. This is consistent with the observation in Figure~\ref{fig:distri_sigma_nonlin} where the distribution of stress is narrower at higher applied stress compared with the lower stress case. The two models (SC and FEM) provide qualitatively similar predictions, although the degree of heterogeneity in FEM is smaller than that in SC. This implies that spatial field correlations tend to reduce the degree of heterogeneity - similar with non-local effects in elasticity \cite{aifantis_gradient_1999}. The data is significantly above the affine limit in all cases.\\
To quantify the evolution of stress heterogeneity during creep, we plot in Fig.~\ref{fig:fraction_total_load_phi0p2} the variation of the fraction of the total load carried by the most loaded sub-domains that represent 20\% of the total material volume \textit{vs.} the normalized compliance, $J/J_0$, as predicted by the SC and FEM models. At the onset of creep ($J/J_0 = 1$), stress is homogeneous and 20\% of the domain volume carries 20\% of the applied load. The degree of heterogeneity increases as creep proceeds and eventually reaches a plateau. In the long time limit stress localization becomes quite pronounced. 
%

%
\section{Conclusion}
The behavior of amorphous polymers in the vicinity of the glass-transition temperature under weakly nonlinear creep is investigated. We first show that the experimental data are well described by the theoretical framework proposed by D. Long and collaborators \cite{long_dynamics_2018,merlette_eyring_2025}, according to which the local relaxation rate is accelerated by stress as $e^{-(\sigma/y)^{2}}$. Both the SC and FEM models reproduce the experimental observation that the macroscopic acceleration follows a similar functional form, namely $e^{-(\Sigma/Y)^{2}}$, with an effective parameter $Y$ that decreases as the compliance $J$ increases.\\
The fact that the macroscopic exponent coincides with its microscopic counterpart in creep---in contrast to step-strain experiments---suggests that creep corresponds to a loading condition in which the stress field remains comparatively homogeneous. Nevertheless, our study shows that stress heterogeneity develops during creep, with both the FEM and SC models predicting the emergence of a substructure that carries an increasing fraction of the applied load. The stress-accelerated relaxation law has a homogenizing effect on the stress field by reducing the lifetime of the slowest domains. As a result, the stress heterogeneity, measured at a fixed strain, decreases with increasing applied stress and, at a fixed applied stress, increases over time before reaching a steady state.\\
Despite the fact that the FEM model explicitly accounts for spatial field-mediated interactions between sub-domains whereas the SC model does not, both approaches yield qualitatively similar predictions that are consistent with the experimental observations. This suggests that, in the specific case of weakly nonlinear creep, spatial correlations are not essential for capturing the qualitative behavior of the system. Overall, creep experiments performed in the weakly nonlinear regime emerge as a robust approach for estimating the local stress-induced acceleration of relaxation in glassy polymer systems.\\

\section*{Acknowledgments}

F. Lequeux and A. Chateauminois wish to express their gratitude to their late friend and colleague, H\'el\`ene Montes, who enthusiastically contributed to the early stages of this work before her passing in November 2023.

\section*{Declarations}

\textbf{Data Availability Statement} The data that support the
findings of this study are available from the corresponding
author upon reasonable request.\\
\textbf{Funding and or Conflict of Interests/Competing Interests} The authors do not have any conflict of interest or competing interest nor funding sources to reports.\\
\textbf{Authors contributions} All the authors contributed equally to this work.
%
%
\begin{appendices}
	\section{Two states model, 2S}
	\label{appendix:appendixA}
	In this model, each domain can either be either in an unrelaxed, elastic deformation state (with a constant modulus $G_i$), or in a fully relaxed state with zero stress. The volume fraction of relaxed states $\phi(t)$ may be related to creep through a derivation based on Eshelby's model~\cite{eshelby1957} - in practice we use Palierne's mean-field expression \cite{palierne_linear_1990} that applies to incompressible bodies. As detailed in \cite{roman-faure_weak_2025}, the compliance may be expressed as a function of the volume fraction of relaxed domains as
	\begin{equation}
		J(t)=J_0 \frac{1+\frac{2}{3}\phi(t)}{1-\phi(t)}.
	\end{equation}
	The relaxation of an increasing fraction of domains during creep leads to an increase in the stress sustained by the remaining unrelaxed domains. Averaging the applied stress $\Sigma$ over the unrelaxed domains yields $\sigma(t)(1- \phi(t)) = \Sigma$, where
	the local stress $\sigma$ is sustained by the unrelaxed domains. This leads to following relation between the local stress and the creep compliance
	\begin{equation}
		\sigma(t)=\frac{2+3\frac{J(t)}{J_0}}{5} \Sigma.
		\label{eq:sigma_two-states}
	\end{equation}
	Taking into account the theoretical value of the acceleration predicted in reference~\cite{conca_acceleration_2017}, the model provides the expression of the macroscopic acceleration factor
	
	\begin{equation}
		F_{cav}=e^{-\left(\frac{2+3\frac{J(t)}{J_0}}{5}\frac{\Sigma}{Y}\right)^2},
		\label{eq:Fcav}
	\end{equation}
	which can be tested against the experimental creep compliance. In this relation, $Y$ is a typical yield stress which was estimated to $Y=14$~\si{\mega\pascal} in reference~\cite{roman-faure_weak_2025}. 
	In the main text we use a quantity representing the fraction $\phi_P$ of the total load carried by a volume fraction $\phi_v$ of the most loaded domains. In the 2S approximation, all domains that are still loaded sustain a stress given by Eq.~(\ref{eq:sigma_two-states}). A simple calculation then show that $\phi_P$ reads 
	\begin{equation}
		\phi_P=\phi_v \frac{3J(t)+2J_0}{5J_0 }.
		\label{eq:psi}
	\end{equation}  
	%
	\section{Elastic inclusion problem}
	\label{appendix:appendixB}
	We derive here the equations for a single elastic spherical inclusion embedded in an infinite elastic matrix. From the seminal work of Eshelby~\cite{eshelby1957}, stress and strain within the inclusion are known to be homogeneous. According to Eqs~(1.13), (1.45), (1.46) and (1.49) in reference~\cite{palierne_linear_1990}, the stress $\sigma_i$ in the inclusion can be written as
	\begin{equation}
		\sigma_i=\frac{5G_iG_m}{2G_i+3G_m}\Gamma,
		\label{eq:sigma_i}
	\end{equation}
	where $G_i$ and $G_m$ are respectively the shear modulus of the inclusion and the matrix and $\Gamma$ is the uniform strain applied to the matrix at infinity.\\
	Introducing  $\gamma_i$, the shear strain inside the inclusion, we can write  $G_i=\sigma_i/\gamma_i$. Eq.~(\ref{eq:sigma_i}) can be thus rewritten in a more general form, as function of the strain and the stress in the inclusion
	\begin{equation}
		\sigma_i=G_m\Gamma+\frac{3}{2}G_m\left(\Gamma-\gamma_i\right).
		\label{eq:sigma_i1}
	\end{equation}
	The above equation is valid whatever the relation between $\sigma_i$ and $\gamma_i$. Inclusion modifies the stress in the matrix and the macroscopic shear modulus becomes (Eqs.~(2.5-2.7) in~\cite{palierne_linear_1990}) 
	\begin{equation}
		G_{MAC}=G_m\left(1+\sum_i f_i\frac{5(G_i-G_m)}{2G_i+3G_m}\right).
		\label{eq:macroscopic modulus}
	\end{equation}
	where $G_{MAC}$ is the macroscopic shear modulus, $G_m$ the shear modulus of the matrix and $f_i$ is the volume fraction of inclusions of type $i$. Relation~(\ref{eq:macroscopic modulus}) is valid only when the total volume fraction of inclusions is small, \textit{i.e.} $\sum_i f_i \ll 1$, as noted, for instance, in reference~\cite{palierne_linear_1990}. Let us now consider the case where the inclusions, with volume fractions $\phi_i$, entirely tile the system, i.e.\ $\sum_i \phi_i = 1$. In the self-consistent approximation, this system is assumed to behave as a homogeneous material with modulus $G_M$. Let us randomly embed in this material inclusions of moduli $G_i$, each with a volume fraction $f_i = \epsilon \phi_i$, where $\epsilon \ll 1$. Since the embedded inclusions are identical to those that make up the matrix, adding them must not change its elastic modulus. Hence, the modulus $G_M$ of the matrix must remain invariant under this operation. This leads to the equation 
	\begin{equation}
		G_M=G_M\left(1+\sum_i \epsilon \phi_i\frac{5(G_i-G_M)}{2G_i+3G_M}\right).
		\label{eq:self consistent1}
	\end{equation}
	which applies for any $\epsilon$. Then, 
	\begin{equation}
		\sum_i \phi_i\frac{(G_i-G_M)}{2G_i+3G_M}=0.
		\label{eq:self consistent2}
	\end{equation}
	This equation gives the value of the macroscopic matrix  modulus $G_M$ which has been successfully used to describe the modulus of blends~\cite{Shi2013}. 
	%
	\section{Self-consistent equations for Maxwell-like viscoelastic inclusions}
	\label{appendix:appendixC}
	In this appendix, we derive the relation between local and macroscopic quantities at a given time $t$ for a material entirely tiled with Maxwell-like viscoelastic inclusions. From the definition of the matrix stress in Eq.~(\ref{eq:matrix_leastoplastic}), we define a new reference frame for the strain, where all the strain components are shifted by $-\Gamma^a$. In this new frame, the strain in the inclusions and in the matrix (labeled by the exponent "new") becomes 
	\begin{equation}
		\Gamma^{new}=\Gamma-\Gamma^a=\Gamma^e,
	\end{equation}
	and
	\begin{equation}
		\gamma^{new}_i=\gamma_i-\Gamma^a.
	\end{equation}
	In the new frame, the matrix is elastic. Thus, the mechanical equilibrium equation between matrix and inclusions (Eq.~(\ref{eq:sigma_i1})) can be applied in the new frame and become  
	\begin{equation}
		\sigma_i=G_m\Gamma^{new}+\frac{3}{2}G_m\left(\Gamma^{new}-\gamma^{new}_i\right).
		\label{eq:sigma_i2}
	\end{equation}
	Introducing the fact that the inclusion is \textit{Maxwell-like} and using Eq.~(\ref{eq:maxwell_inclusion_1}) and $\Sigma=G_m\Gamma^{new}$ leads to
	\begin{equation}
		5\Sigma-\left(3G_m+2G_i\right)\gamma_i^e+3 G_m\left(\Gamma^a-\gamma_i^a\right)=0
		\label{eq:gamma_i}
	\end{equation}
	Its time derivative may be written using Eqs.~(\ref{eq:maxwell_inclusion_1}) and (\ref{eq:maxwell_inclusion_2}) as
	\begin{equation}
		5\frac{\partial \Sigma}{\partial t}-\frac{3G_m+2G_i}{G_i}\frac{\partial \sigma_i}{\partial t}+3 G_m\left( \frac{\partial \Gamma^a}{\partial t}-\frac{\sigma_i}{G_i\tau_i}\right)=0
		\label{eq:time_derivative}
	\end{equation}
	This equation gives the stress evolution of a Maxwell-like inclusion $i$ embedded in a viscoelastic matrix of modulus $G_m$ of macroscopic stress $\Sigma$ and of anelastic deformation $\Gamma_a$.
	%
	\section{Self-consistent equations}
	\label{appendix:appendixD}
	In order to derive the expressions of the time-dependent stress within inclusions, we will first show that Eq.~(\ref{eq:self consistent2}) applies even in the presence of Maxwell inclusions. Because the inclusions are randomly distributed and tile entirely the material - and thus $\sum_i \phi_i=1$ - any flat surface that cuts the material exhibits a random distribution of forces which is representative of the bulk. As a consequence, it sustains a force which is the average of local forces. Within the framework of this self-consistent approximation, the macroscopic stress $\Sigma$ is thus equal to the average of the local stresses $\sigma_i$. This leads to the equation
	\begin{equation}
		\Sigma=\sum_i \phi_i \sigma_i 
		\label{eq:sigma conservation}
	\end{equation}
	A similar relation may be written for the strain: 
	\begin{equation}
		\Gamma=\sum_i \phi_i\gamma_i.
		\label{eq:Gamma}
	\end{equation}
	At given time $t$, we can apply a virtual instantaneous increase in the macroscopic stress $\delta\Sigma=G_m \delta\Gamma$, while keeping  constant the anelastic deformations in the matrix and in the inclusions. This leads to a virtual increase in the elastic deformation of inclusions, which can be obtained just by differentiating Eq.~(\ref{eq:gamma_i})
	\begin{equation}
		\delta\gamma_i=\frac{5\delta\Sigma}{3G_m+2G_i}
		\label{eq:virtual_gamma_i}
	\end{equation}
	Writing $\delta\Gamma=\sum_i \phi_i\delta\gamma_i$ from the conservation equation~(\ref{eq:Gamma}), we get
	\begin{equation}
		\sum_i \phi_i\frac{(G_i-G_m)}{2G_i+3G_m}=0,
		\label{eq:self consistent3}
	\end{equation}
	an equation identical to equation~(\ref{eq:self consistent2}). Hence, the macroscopic modulus is the same whatever the extent of the plastic deformations within the inclusions. 
	In the particular case where all inclusions have the same elastic modulus $G_i=G_0$, this leads to $G_m=G_0$.\\
	
	We focus now on creep under a constant applied stress which yields $d\Sigma/dt=0$. In that case, $\partial \Gamma/\partial t= \partial \Gamma^a /\partial t$. Thus, using equations~(\ref{eq:Gamma}), (\ref{eq:maxwell_inclusion_0}) and (\ref{eq:maxwell_inclusion_2}) we get 
	\begin{equation}
		\frac{\partial\Gamma^a}{\partial t}=\sum_j\left(\frac{\phi_j}{G_j}(\frac{\partial\sigma_j}{\partial t}+\frac{\sigma_j}{\tau_j}) \right)
		\label{eq:Gamma_p}
	\end{equation}
	Equation~(\ref{eq:time_derivative}) which couples the time evolution of inclusions with that of the matrix, can now be written only in terms of the inclusions stress
	\begin{equation}
		\frac{\partial \sigma_i}{\partial t}=\frac{3 G_m G_i}{3G_m+2G_i}\left(\sum_j\left(\frac{\phi_j}{G_j}\left(\frac{\partial\sigma_j}{\partial t}+\frac{\sigma_j}{\tau_j}\right) \right)-\frac{\sigma_i}{G_i\tau_i}\right)
		\label{eq:evol_sigma}
	\end{equation}
	Equation~(\ref{eq:evol_sigma}) needs to be integrated with initial conditions (according to Eq.~(\ref{eq:sigma_i})):
	\begin{equation}
		\sigma_i(t=0)=\frac{5G_i }{2G_i+3G_m}\Sigma
	\end{equation}
	The dashpot strain of the Maxwell model may be calculated using eqn~(\ref{eq:maxwell_inclusion_2}).\\
	
	In what follows, we set $G_i=G_0$ for all the Maxwell elements. From Eq.~(\ref{eq:self consistent3}), we get $G_m=G_0$. In that case the term 
	$\sum_j\frac{\phi_j}{G_0}\frac{\partial\sigma_j}{\partial t}=0$ because the total stress is constant, and Eq.~(\ref{eq:evol_sigma}) reduces to 
	\begin{equation}
		\frac{\partial \sigma_i}{\partial t}=M_{ij} \sigma_j,
		\label{eq:system}
	\end{equation}
	where
	\begin{equation} 
		M_{ij}=\frac{3}{5}\left(\frac{\phi_j}{\tau_j}-\frac{\delta_{ij}}{\tau_i}\right).
		\label{eq:matrixM1}
	\end{equation}
	Note that in the trivial case where all inclusions are identical the matrix $M_{ij}$ reduces to a scalar, $M_{11}=0$. The stress in the inclusion remains constant, and we recover that the macroscopic behavior is that of a single inclusion. 
	%
	\section{Numerical implementation}
	\label{appendix:appendixE}
	Direct numerical calculations of $J(t)$ using Eqs.~(\ref{eq:system}) and (\ref{eq:matrixM1}) become difficult since some domains exhibit a local relaxation time that may become smaller than the time step. In order to circumvent this difficulty, the following scheme is implemented starting from Eq.~(\ref{eq:system}) where we have substituted Eq.~(\ref{eq:matrixM1}) for $M_{ij}$. We solve Eq.~(\ref{eq:matrixM1}) using the elastic strains $\gamma_i^e$ instead of the local stress based on relation~(\ref{eq:maxwell_inclusion_1})
	\begin{equation}
		\frac{\partial \gamma_i^e}{\partial t}=\frac{3}{5} \left[\sum_j\frac{\phi_j}{\tau_j}\gamma_j^e-\frac{\gamma_i^e}{\tau_i}\right].
		\label{eq:equA1}
	\end{equation}
	If the time step of the iterative calculation is selected so that the first term in the right hand side of the above equation can be considered as a constant during each iteration, then Eq.~(\ref{eq:equA1}) can be rewritten as
	\begin{equation}
		\frac{\partial \gamma_i^e(t)}{\partial t}=\frac{3}{5} \left[k(t)-\frac{\gamma_i^e(t)}{\tau_i}\right],
	\end{equation}
	with
	\begin{equation}
		k(t)=\sum_j\frac{\phi_j}{\tau_j}\gamma_j^e(t).
	\end{equation}
	Accordingly, 
	\begin{equation}
		\frac{\partial}{\partial t}\left(\gamma_i^e -\tau_i k(t)\right)=-\frac{3}{5\tau_i}\left(\gamma_i^e-\tau_i k(t)\right),
	\end{equation}
	and the elastic deformation term $\gamma_i^e(t)$ at time $t+dt$ can thus be written as 
	\begin{equation}
		\gamma_i^e(t+dt)= \gamma_i^e(t)e^{-\frac{3dt}{5\tau_i}}+\tau_i k(t)\left(1-e^{-\frac{3 dt}{5 \tau_i}}\right),
	\end{equation}
	where the time step $dt$ of each iteration is selected such to fulfill the criterion
	\begin{equation}
		|(k(t+dt)-k(t))|<  \max [1/\tau_i \left( \gamma_i^e(t+dt)-\gamma_i^e(t) \right)] \zeta,
	\end{equation}
	where we take $\zeta=10^{-4}$.
	
\end{appendices}



\clearpage

\section*{ SI 1. Self consistent modeling of the linear viscoelastic behaviour of PEI in the glass transition range}
The self-consistent model (SC) is used to compute the linear viscoelastic modulus $G^*$ over an extended range of frequencies and predictions are compared to the corresponding experimental curve. As detailed by Shi~\textit{et al.}~[28], the shear viscoelastic modulus in the linear regime is given by the following integral equation 
\begin{equation}
	\int \frac{2G^*_{max}(\omega,\tau)-2G^*(\omega,\tau)}{2G^*_{max}(\omega,\tau)+3G^*(\omega,\tau)}P(\tau)d\tau=0
\end{equation}
where $G^*_{max}(\omega,\tau)$ is the value of the viscoelastic modulus of a single Maxwell element. In order to avoid numerical difficulties in solving this differential equation, the calculation was carried out using a recursive approach which consists in iterating the following suite, starting from any value of G*: 
\begin{equation}
	G^*_{n+1}=\frac{1}{\int\frac{5 P(\tau) d\tau}{2*G^*_i+3G^*_n}},
	\label{eq:Gstar}
\end{equation}
where $P(\tau)$ is the relaxation time distribution.
\begin{figure} [!htp]
	\centering
	\includegraphics[width=1\linewidth]{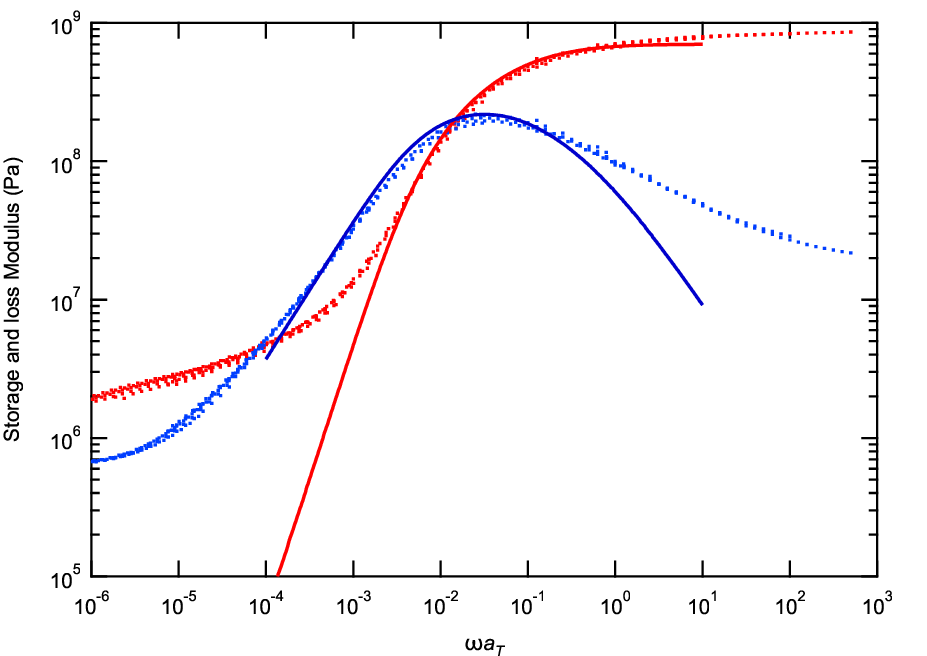}
	\caption{Master curve giving the storage,$G'$, (red) and loss, $G''$, (blue)  shear linear viscoelastic modulus of PEI as a function of reduced frequency at $T_{ref}=T_g=213$~\si{\celsius}. Dots: experimental values; solid lines: theoretical values calculated using Eqn.~\ref{eq:Gstar} ($s=1.86$ and $\tau_0=0.21$~\si{seconds}).}
	\label{fig:mastercurve_G}
\end{figure}
The resulting master curve is shown in Fig.~\ref{fig:mastercurve_G} together with experimental data for the PEI polymer at $T_{ref}=$213~\si{\celsius}. A good agreement is observed between theoretical and experimental data in the vicinity of the glass transition which corresponds to the investigated creep regime.\\

\section*{SI 2. Effect of the width of the relaxation time distribution on creep acceleration.}
\begin{figure} [!ht]
	\centering
	\includegraphics[width=1\linewidth]{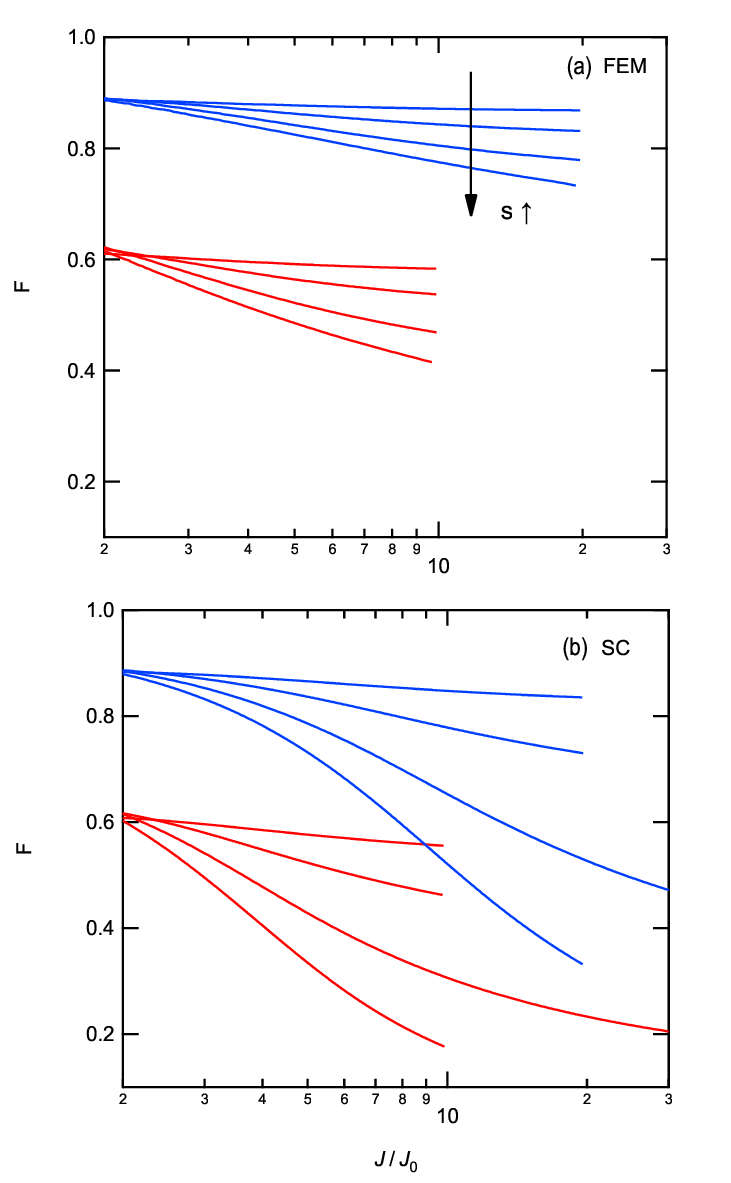}
	\caption{Acceleration factor as a function of the normalized compliance $J/J_0$ for different widths of the relaxation time distribution and two values of the applied stress. (a) FEM simulations, (b) SC model. Blue : $\Sigma=5$~\si{\mega\pascal}, red: $ \Sigma=10$~\si{\mega\pascal}. From top to bottom, the width of the distribution is $s=$~0.5, 1, 1.86 and 3.}
	\label{fig:SI2_fig}
\end{figure}

We consider here the acceleration of creep in the non linear regime for relaxation time distributions with increasing width (from $s=0.5$ to 3), \textit{i.e.} glassy materials with increasing disorder. The FEM and SC results reported in Fig.~\ref{fig:SI2_fig} for two values of the applied stress $\Sigma$ show that the decrease of the acceleration factor $F$ with increasing normalized compliance $J/J_0$ is enhanced when the width of the distribution increases. In other words, the effect of material disorder is to enhance creep acceleration, an effect which is captured in the same qualitative way by both SC and FEM models.\\

\end{document}